\documentstyle[12pt]{article}

\begin{document}

\title{
 TIME-SYMMETRIZED COUNTERFACTUALS IN QUANTUM THEORY}
\author{ Lev Vaidman}
\date{}
\maketitle

\begin{center}
{\small \em School of Physics and Astronomy \\
Raymond and Beverly Sackler Faculty of Exact Sciences \\
Tel Aviv University, Tel-Aviv 69978, Israel. \\}
\end{center}

\vspace{2cm}
\begin{abstract}
  Recently, several authors have criticized the time-symmetrized quantum
  theory originated by the work of Aharonov et al. (1964).  The core
  of this criticism was a proof,  appearing in various forms,
  which showed that the counterfactual interpretation of time-symmetrized
  quantum theory cannot be reconciled with standard quantum
  theory.  I argue here that the apparent contradiction is due to
  a logical error. I analyze the concept of counterfactuals in quantum
  theory and introduce time-symmetrized counterfactuals. These
  counterfactuals do not lead to any contradiction with the predictions of
  quantum theory. I discuss applications of time-symmetrized
  counterfactuals to several surprising examples and show the usefulness
  of the time-symmetrized quantum formalism.
\end{abstract}

\vfill\break

 \noindent
{\bf 1. Introduction.~~ }
I shall discuss measurements performed on  a pre- and post-selected quantum system, i.e.
 at a time between two other measurements.
The time-symmetrized formalism  describing  such systems was proposed
by Aharonov, Bergmann, and Lebowitz (ABL) (1964) and has been developed in
recent years. A partial list of references includes Aharonov et al.
(1985), Aharonov and Vaidman (1990, 1991).  Several authors criticized
the time-symmetrized approach to quantum theory in general and some of its
particular applications. The most representative example is the work
of Sharp and Shanks (1993). They presented a proof, which was later
repeated and used by others, that the counterfactual interpretation of the
ABL probability rule (Eq. 2 below) cannot be reconciled with
the standard quantum theory. I shall claim here that the proof
contains a logical
error. I shall analyze the concept of counterfactuals in quantum theory
and introduce time-symmetrized counterfactuals. The ABL rule can be
applied to such counterfactuals and I shall present situations for
which it is useful.

The plan of this work is as follows.  In Section 2 I present a brief
review of the time-symmetrized formalism.  In section 3 I analyze the
concept of counterfactuals in quantum theory and introduce
time-symmetrized counterfactuals.  Section 4 is devoted to the analysis
of the inconsistency proof of Sharp and Shanks and its variations.  In
section 5 I discuss related time asymmetry preconceptions in quantum
theory. In section 6 the time symmetry (and asymmetry) of the process
of quantum measurement is analyzed in order to give a rigorous context
to the previous discussion of the time symmetry of the ABL rule.
Applications of  time asymmetric and time symmetric  counterfactuals in
quantum theory are 
considered in sections 7-9. Section 10 summarize the arguments of the
paper.

\vskip 1.cm \noindent
{\bf  2. Time-Symmetrized Formalism.~~}
In standard quantum theory a complete description of a system at a
given time is given by a quantum state $|\Psi \rangle$. It yields the
probabilities for all outcomes $a_i$ of a measurement of any observable 
$A$ according to the equation
\begin{equation}
  \label{prob1}
  {\rm Prob}(a_i) = |\langle \Psi | {\bf P}_{A=a_i} | \Psi \rangle | ,
\end{equation}
where ${\bf P}_{A=a_i}$ is the projection operator on the subspace defined
by $A= a_i$.  Eq. 1 is intrinsically asymmetric in time: the state
$|\Psi\rangle$ is determined by some measurements in the past and it
evolves toward the future.  The time evolution between the
measurements, however, is considered time symmetric since it is
governed by the Schr\"odinger equation for which each forward evolving
solution has its counterpart (its complex conjugate with some other
well understood simple changes) evolving backward in time. The
asymmetry in time of the standard quantum formalism is manifested in the
absence of  the quantum state evolving backward in time from 
future measurements (relative to the time in question).

Time-symmetrized quantum theory describes a system at a given time by
a two-state vector $\langle \Psi_2| |\Psi_1 \rangle$. It yields the
(conditional) probabilities for all outcomes $a_i$ of a measurement of
any observable $A$ according to the generalization of the ABL formula
(Aharonov and Vaidman, 1991):
\begin{equation}
  \label{ABL}
 {\rm Prob}(a_i) = {{|\langle \Psi_2 | {\bf P}_{A=a_i} | \Psi_1 \rangle |^2}
\over{\sum_j|\langle \Psi_2 | {\bf P}_{A=a_j} | \Psi_1 \rangle |^2}} .
\end{equation}

 The time symmetry means that $\langle \Psi_2|$ and $|\Psi_1
\rangle$ enter the equations, and thus govern the observable results,
on equal footings.
Moreover, the time symmetry  means that, in regard to time symmetric
measurements, a system described by the
two-state vector $\langle \Psi_2| |\Psi_1 \rangle$ is identical to a system described by the
two-state vector $\langle \Psi_1| |\Psi_2 \rangle$.
I will analyze the time symmetry of the process of measurement in
section 6; here I only point out that ideal measurements are time
symmetric. Indeed, the  symmetry under the interchange of $\langle
\Psi_2|$ and  $|\Psi_1 \rangle$ is explicit in Eq. 2 which refers to
ideal measurements.

Another basic concept of time-symmetrized two-state vector formalism is
{\em weak value}. An  (almost) standard measurement
procedure  for measuring observable $A$ with weakened coupling (which we call {\em weak
  measurement}, Aharonov and Vaidman 1990) yields  the {\em weak
  value} of $A$:
\begin{equation}
A_w \equiv { \langle{\Psi_2} \vert A \vert\Psi_1\rangle 
\over \langle{\Psi_2}\vert{\Psi_1}\rangle } .
\label{wv}
\end{equation}
Here again, $\langle \Psi_2|$ and $|\Psi_1
\rangle$ enter the equations
on equal footings. However, when we interchange $\langle \Psi_2|$ and $|\Psi_1 \rangle$, the weak
value changes to its complex conjugate. Thus, in this situation, as for
the Schr\"odinger equation, time reversal is accompanied by complex
conjugation.

In order to explain how to obtain a quantum system described at a
given time $t$ by a two-state vector $\langle \Psi_2| |\Psi_1 \rangle$
we shall assume for simplicity that the free Hamiltonian of the system
is zero. In this case, it is enough to prepare the system at time
$t_1$ prior to  time $t$ in the state $|\Psi_1\rangle$, to ensure
no disturbance between $t_1$ and $t$ as well as between $t$ and $t_2$,
and to find the system at $t_2$ in the state $|\Psi_2\rangle$. It is
crucial that $t_1 < t < t_2$, but the relation between these times and
``now'' is not fixed. The times $t_1, t, t_2$ might all be in the past, or we
can discuss future measurements and then they are all in the
future; we just have to agree to discard all cases when the
measurements at  time $t_2$ does not yield the outcome
corresponding to the state $|\Psi_2\rangle$.

Note the asymmetry between the measurement at $t_1$ and the
measurement at $t_2$. Given an ensemble of quantum systems, it is
always possible to prepare all of them in a particular state
$|\Psi_1\rangle$, but we cannot ensure finding the system in a
particular state $|\Psi_2\rangle$.  Indeed, if the pre-selection
measurement yielded a result different from projection on
$|\Psi_1\rangle$ we can always change the state to $|\Psi_1\rangle$,
but if the measurement at $t_2$ did not show $|\Psi_2\rangle$, our
only choice is to discard such a system from the ensemble.
 This  asymmetry,
however, is not relevant to the problem we consider here.  We study
the symmetry relative to the measurements at time $t$ for a {\em given} pre-
and post-selected system, and we do not investigate the time-symmetry
of obtaining such a system. The only important detail is that the
interaction  at time $t$ has to be time symmetric.
 See more discussion below, in section 6.

\vskip 1.cm \noindent
{\bf 3. Counterfactuals.~~}
 A general form of a counterfactual statement
is
\begin{quotation}
 { \bf (i)}
{\em If it were that $\cal A$, then it would be that $\cal B$.}
\end{quotation}
There are  many philosophical discussions on the concept of
counterfactuals and especially on  time's arrow in counterfactuals.
Many of the discussions, e.g. Lewis (1986), Bennett (1984), are
related to $\cal A$: How come $\cal A$ if in the actual world $\cal A$ is
not true? Do we need a miracle (a violation of the fundamental law
of nature) for $\cal A$?  Does $\cal A$ come
by itself, or it is accompanied by other changes? However, these
questions are not relevant to the problem of counterfactuals in
quantum theory.  The questions about $\cal A$ are not relevant because
$\cal A$ depends solely on an external system which is not under
discussion by the definition of the problem.  Indeed, in quantum theory
the counterfactuals have a very specific form:\footnote{This
  definition of counterfactuals in quantum theory is broad enough for
  discussing issues relevant to this paper. However, in some cases the
  term ``counterfactuals'' was used differently. For example, in
  Penrose (1994, p.240) ``{\em counterfactuals} are things that might
  have happened, although they did not in fact happen.''}
\begin{quotation}

 $\cal A$ =  a measurement  $\cal M$   is performed

 $\cal B$ = the outcome of $\cal M$  has  property $\cal P$
\end{quotation}
The measurement $\cal M$  might consist of
measurements of several observables  performed together. The
property $\cal P$ might be a certain
relation between the results of measurements of these observables or a
probability for a certain relation or for a certain outcome.

It is assumed that the experimenter can make any decision about which
measurement to perform and the question how he makes this decision is
not considered. It is assumed that the experimenter and his measuring
devices are not correlated in any way with the state of the system
prior to the measurement. Thus, in the world of the quantum system no
miracles are needed and no changes relative to the actual world have to be
made for different $\cal A$'s.\footnote 
  {Indeterminism of standard
  quantum theory allows us to discuss even the worlds which include the
  experimenter without invoking miracles. Consider an experimenter who
  chooses between different measurements according to a random outcome
  of another quantum experiment.}

Although one can define counterfactuals of this form in the framework
of classical theory, they are of no interest because they are
equivalent to some ``factual'' statements. In classical physics any
observable has always a definite value and a measurement of the observable
yields this value. Therefore, we can make a one to one correspondence
between ``the outcome of a measurement of an observable $C$ is $c_i$''
and ``the value of $C$ is $c_i$''. The latter is independent of
whether the measurement of $C$ has been performed or not and,
therefore, statements which are formally counterfactual about results
of possible measurements can be replaced by ``factual''
(unconditional) statements about values of corresponding observables.
In contrast, in standard quantum theory, observables, in general, do not
have definite values and therefore we cannot always reduce the above
counterfactual statements to ``factual'' statements.

Most of the discussions of counterfactuals in quantum theory are in the
context of EPR-Bell type experiments. Some of the examples are Skyrms
(1982), Peres (1993), Mermin (1989) (which, however, does not use the
word counterfactual), and Bedford and Stapp (1995) who even present an
analysis of a Bell-type argument in the formal language of the Lewis
(1973) theory of counterfactuals. The common situation is that a
composite system is described at a certain time by some entangled
state and then an array of incompatible measurements on this system at
a later time is considered. Various conclusions are derived from
statements about the results of these measurements. Since these
measurements are incompatible they cannot be all performed together,
so it must be that at least some of them were not actually performed.
This is why they are called counterfactual statements.

These counterfactuals are explicitly asymmetric in time. The
asymmetry is neither in $\cal A$ nor in $\cal B$; both are about a
single time  $t$.
 The asymmetry is in the description of the actual world.
 The {\it past} and not the {\it
  future}  (relative to $t$) of a system is given. 

This, however, is not the only asymmetry of the counterfactuals in
quantum theory as they are usually considered. A different asymmetry
(although it looks very similar) is in what we assume to be ``fixed'',
i.e.,  which properties of actual world we assume to be true in
possible counterfactual worlds.  The {\it past} and not the {\it
  future} of the system is fixed.

It seems that while the first asymmetry can be easily removed, the
second asymmetry is unavoidable. According to standard quantum theory
a system is described by its quantum state. In the actual world, in
which a certain measurement has been performed at time $t$ (or no
measurement has been performed at $t$) the system is described by a
certain state before $t$, and by some state after time $t$. In the
counterfactual world in which a different measurement was performed at
time $t$, the state before $t$ is, of course, the same, but the state
after time $t$ is invariably different (if the observables measured in
actual and counterfactual worlds have different eigenstates).
Therefore, we cannot hold fixed the quantum state of the system in the
future.\footnote {Note that none of these asymmetries exists in
  classical case because when a complete description of a classical
  system is given at one time, it yields and fixes the complete
  description at all times and (ideal) measurements at time $t$ do not
  change the state of a classical system.}

The argument above shows that for  constructing  time
symmetric counterfactuals we have to give up the description of a
quantum system by its quantum state. Fortunately we can do that without loosing
anything except the change due to the measurement at time $t$ which
caused the difficulty. A quantum state at a given time is completely
defined by the results of a complete set of measurements performed
prior to this time.  Therefore, we can take the set of all results
performed on a quantum system as a description of the world of the
system instead of describing the system by its quantum state. (This
proposal will also help to avoid ambiguity and some controversies
related to the description of a single quantum system by its quantum
state.) Thus, I propose the following  definition of
counterfactuals in the framework of quantum theory:

\begin{quotation}
{ \bf (ii)}
{\em  If  a measurement ${\cal M}$  were performed at
  time $t$, then it would have property ${\cal P}$, provided  that the results of all measurements
performed on the system at all times except the time $t$ are fixed.}
\end{quotation}
 
For time asymmetric situations in which only the results of
measurements performed before $t$ are given (and thus only these
results are fixed) this definition of counterfactuals is equivalent to
the counterfactuals as they usually have been used. However, when the
results of measurements performed on the systems both before and after
the time $t$ are given, definition (ii) yields novel time-symmetrized
counterfactuals. In particular, for the ABL case, in which {\em
  complete} measurements are performed on the system at $t_1$ and
$t_2$, $t_1 < t <t_2$, we obtain
\begin{quotation}
{ \bf (iii)}
{\em  If  a measurement of an observable $C$ were performed at
  time $t$, then  the probability for
$C=c_i$ would equal $p_i$, provided  that the results of  measurements
performed on the system at times $t_1$ and $t_2$  are fixed.}
\end{quotation}
The ABL formula (2) yields correct probabilities for counterfactuals
defined as in (iii), i.e., in the experiment in which $C$ is measured at
time $t$  on the systems from pre- and
post-selected ensemble defined by fixed outcomes of the measurements at
$t_1$ and $t_2$ (all such   systems and only such systems are
considered)
the frequency of an outcome $c_i$ is  $p_i$.

 For  the ABL situation  one
can also define a time
{\em asymmetric} counterfactual:
\begin{quotation}
{\bf (iv)} Given the results of measurements at $t_1$ and $t_2$, $t_1 < t
<t_2$ (in the actual world),
if  a measurement of a  observable $C$ were performed at
  time $t$, then  the probability for
$C=c_i$ would equal $p_i$, provided  that the results of all measurements
performed on the system at all times before  time $t$ are fixed.
\end{quotation}
In the framework of standard quantum theory the information about the result
of measurement at $t_2$ is irrelevant: the probability for
$C=c_i$ does not depend on this result. Thus, it is obvious that the
ABL formula (2), which includes the result of the measurement at time
$t_2$ explicitly,  does not yield counterfactual probabilities according
to definition (iv).

 One might modify definition (iv) in the framework of
some ``hidden variable'' theory with a natural additional requirement of
fixing the hidden variables of the system in the past. The properties of
such counterfactuals will depend crucially on the details of the hidden
variable theory (see the discussion of Aharonov and Albert (1987) in the
framework of Bohm's theory), but  the ABL formula (2) is not valid
for any such modification. 
In order to show this consider a spin-${1\over 2}$ particle which  was found  at
$t_1$ and  at  $t_2$ in the same state
$|{\uparrow}_z\rangle$
 (and no measurement has been performed at $t$). 
 We ask what is the (counterfactual) probability for finding
spin ``up'' in the direction $\hat \xi$ which makes an angle $\theta$
with the direction $\hat z$, at the intermediate time $t$. In this case,
 hidden variables, even if  they exist, cannot change that probability
because any   particle found at $t_1$ in the state $|{\uparrow}_z\rangle$, irrespectively of
its hidden variable, 
yields the outcome ``up'' in the 
measurement at  $t_2$. Therefore, the statistical predictions
about the intermediate measurement at time $t$ must be the same as for
the pre-selected only ensemble (these are {\em identical} ensembles in
this case), i.e.
\begin{equation}
  \label{abl-qm}
  {\rm Prob}({\uparrow}_\xi) =|\langle {\uparrow}_\xi |
{\uparrow}_z\rangle|^2 = \cos^2(\theta/2).
\end{equation}
The ABL formula, however, yields: 
\begin{equation}
  \label{abl-xiz}
  {\rm Prob}({\uparrow}_\xi) = {{|\langle {\uparrow}_z |
      {\bf P}_{{\uparrow}_\xi} | {\uparrow}_z \rangle |^2}\over{|\langle
      {\uparrow}_z | {\bf P}_{{\uparrow}_\xi} | {\uparrow}_z \rangle |^2
    +|\langle {\uparrow}_z | {\bf P}_{\downarrow_\xi} | {\uparrow}_z \rangle
  |^2}}= {{ \cos^4(\theta/2)}\over{ \cos^4(\theta/2) +  \sin^4(\theta/2)}} .
\end{equation}
The fact that the ABL formula (2) does not hold for counterfactuals
defined in (iv) or its modifications is not surprising. Definition
(iv) is explicitly asymmetric in time. The ABL formula, however, is
time symmetric and therefore it can hold only for time-symmetrized
counterfactuals.

A recent study of time's arrow and counterfactuals in the framework of
quantum theory by Price (1996) seems to support my definition (ii).
Let me quote from his section ``Counterfactuals: What should we
fix?'':
\begin{quotation}
Hold fixed the past, and the same difficulties arise all over
again. Hold fixed merely what is accessible, on the other hand, and it
will be difficult to see why this course was not chosen from the
beginning. (1996, 179)
\end{quotation}
This quotation looks very much like my proposal. Indeed, I find
many arguments  in his book pointing in the same direction. However, 
in fact,  this quotation represents a time asymmetry: according to Price
``merely what is accessible''  is ``an accessible
past''. But this is not the time asymmetry of the physical theory;   Price writes: ``no physical asymmetry is
required to explain it.''  Although the book includes an extensive
analysis of a photon passing through two polarizers -- the classic
setup for the ABL case, I found no
explicit discussion of a possible measurement in between, the problem we discuss here.\footnote{Price briefly and critically mentions the ABL paper. He
  writes (1996, 208): ``What they [ABL] fail   to note, however, is
  that  their argument  does nothing to address the problem for those
 who disagree with Einstein -- those  who think that the state
 function is a complete description, so that the change that takes
 place on measurements is a real change in the world, rather than
 merely change in our knowledge of the world.'' This seems to me an
 unfair criticism:  ABL clearly state that in the situations they
 consider ``the complete description'' is given by  {\it two} wave
 functions (see more in Aharonov and Vaidman 1991). Moreover, it seems
 to me that the  development of this time-symmetrized quantum formalism
 is not too far from the spirit of the ``advanced action'' -- the Price
 vision of the solution of the time's arrow problem.}

\vskip 1.cm \noindent
{\bf 4. Inconsistency proofs.~~}
The key point of the criticism of the time-symmetrized quantum theory (Sharp and Shanks 1993; Cohen 1995; Miller 1996)
is the conflict between counterfactual interpretations of the ABL rule
and predictions of quantum theory. I shall argue here that the inconsistency proofs
 are unfounded and therefore the criticism
essentially breaks  apart.

The  structure of all these inconsistency proofs is a follows.
  Three consecutive measurements are considered. The
first is the preparation of the state $|\Psi_1\rangle$ at  time
$t_1$.  The probabilities for the results $a_i$ of the second
measurement at  time $t$ are considered. The final measurement
at  time $t_2$ is introduced in order to allow the analysis 
using the ABL formula.  Sharp and Shanks consider three consecutive
spin component measurements of a spin-${1\over 2}$ particle in different
directions. Cohen analyzes a particular single-particle
interference experiment. It is a variation  of
the Mach-Zehnder interferometer with two detectors for the final
measurement and the possibility of placing  a third detector for the
intermediate measurement. Finally, Miller repeated the argument for a
 system of tandem Mach-Zehnder interferometers.
  In all these cases the ``pre-selection only''
situation is considered.
It is unnatural to apply the time-symmetrized formalism for such
cases. However, it must be possible.  Thus, I need not show that the
time-symmetrized formalism has an advantage over the standard
formalism for describing these situations, but  I  only that it is 
consist  with the predictions of the standard quantum theory. 

In the standard approach to quantum theory the
probability for the result of a measurement of $A$ at  time $t$
is given by Eq. 1. The claim of all the proofs is that the counterfactual
interpretation of the ABL rule yields a different result.  In all cases
the final measurement at time $t_2$ has two possible outcomes
which we signify as ``$1_f$'' and ``$2_f$''.
The suggested application of the ABL rule is as follows. The
probability for the result $a_i$ is:
\begin{equation}
  \label{p1}
  {\rm Prob}(A=a_i)~=~  {\rm Prob}(1_f) ~ {\rm Prob}(A=a_i |1_f) +
  {\rm Prob}(2_f) ~{\rm Prob}(A=a_i |2_f),
\end{equation}
where $ {\rm Prob}(A=a_i |1_f)$ and $ {\rm Prob}(A=a_i |2_f)$ are the 
conditional probabilities given by the ABL formula, Eq. 2, and $ {\rm
  Prob}(1_f)$ and $ {\rm Prob}(2_f)$ are the probabilities for the
results of the final measurement.
In the proofs, the authors show that Eq. 6 is not valid and conclude
that  the ABL formula is not applicable for this example and 
therefore that it is not applicable in general.

I will argue that the error in calculating equality (6) is not in the conditional
probabilities given by the ABL formula, but   in the calculation of the probabilities $ {\rm
  Prob}(1_f)$ and $ {\rm Prob}(2_f)$ of the final measurement. In all three
cases it was calculated on the assumption that {\rm no} measurement
took place at  time $t$. Clearly, one cannot make this assumption
here since then the discussion about the probability of the result of
the measurement at  time $t$ is meaningless. Unperformed
measurements have no results (Peres, 1978). Thus, there is no  surprise that the
value for the probability ${\rm Prob}(A=a_i)$ obtained in this way
comes out different from the value predicted by the quantum theory.

Straightforward calculations show that if one uses the formula (6)
with the probabilities $ {\rm Prob}(1_f)$ and $ {\rm Prob}(2_f)$
calculated on the condition that the intermediate measurement has been
performed, then the outcome is the same as predicted by the standard
formalism of quantum theory.  Consider, for example, the experiment
suggested by Sharp and Shanks, consecutive spin measurements with
the three directions in the same plane and the relative angles
$\theta_{ab}$ and $\theta_{bc}$. The probability for the final result
``up'' is

\begin{equation}
  \label{p1f}
  {\rm Prob}(1_f) =
\cos^2(\theta_{ab}/2)\cos^2(\theta_{bc}/2)+
\sin^2(\theta_{ab}/2)\sin^2(\theta_{bc}/2),
\end{equation}
  and the probability for
the final result ``down'' is
\begin{equation}
  \label{p12}
 {\rm Prob}(2_f)
=\cos^2(\theta_{ab}/2)\sin^2(\theta_{bc}/2)+
\sin^2(\theta_{ab}/2)\cos^2(\theta_{bc}/2). 
\end{equation}
 The ABL formula yields
\begin{equation}
  \label{p1abl}
 {\rm Prob}(up |1_f)={{\cos^2(\theta_{ab}/2)\cos^2(\theta_{bc}/2)}
  \over {\cos^2(\theta_{ab}/2)\cos^2(\theta_{bc}/2)+
    \sin^2(\theta_{ab}/2)\sin^2(\theta_{bc}/2)}}
\end{equation}
 and 
\begin{equation}
  \label{p2abl}
 {\rm
  Prob}(up |2_f)={{\cos^2(\theta_{ab}/2)\sin^2(\theta_{bc}/2)}
  \over {\cos^2(\theta_{ab}/2)\sin^2(\theta_{bc}/2)+
    \sin^2(\theta_{ab}/2)\cos^2(\theta_{bc}/2)}}.
\end{equation}
 Substituting all
these equations into Eq. 6 we obtain
\begin{equation}
  \label{paiabl}
 {\rm
  Prob}(up)=\cos^2(\theta_{ab}/2).
\end{equation}
This result coincide with the prediction of the standard quantum
theory. It is a straightforward exercise to show in the same way that
no inconsistency arises also in the examples of Cohen\footnote{In Cohen's
  example the measurement at time $t_2$ is not a complete measurement
  and therefore the ABL formula (2) is not applicable for this
  case. The analysis requires a generalization of the ABL formula
  given in Vaidman (1997).} and Miller.

I have shown that one can apply the time-symmetrized formalism, including
the ABL formula, for analyzing the examples which allegedly lead to
contradictions in the inconsistency proofs. In my analysis there was
nothing ``counterfactual''. The proofs, however, claimed to show that
a ``counterfactual interpretation'' of the ABL rule leads to
contradiction.
 What I have shown is that the examples
presented in the proofs do not correspond to counterfactual situations
and this is why they cannot be analyzed in a counterfactual way. The
contradictions in the proofs arise from a logical error in taking
together the statement ``no measurement has been performed at $t$''
and a statement about probability of a result of this measurement
which requires ``the measurement has been performed at $t$''.
Let me demonstrate how  similar erroneous ``counterfactual''
reasoning  can  lead to a
contradiction in quantum theory even in cases when the ABL rule is not
involved.  Consider two consecutive measurements of $\sigma_x$
performed on a spin-${1\over 2}$ particle prepared in a state
$|{\uparrow}_z\rangle$. Let us ask (using the language of Sharp and
Shanks) what is the probability that these measurements {\it would
  have had} the results $\sigma_x(t_1) =\sigma_x(t_2) =1$ given that
no such measurements in fact took place. Each spin measurement, if
performed separately, has probability ${1\over 2}$ for the result $\sigma_x
=1$. According to standard quantum theory the fact that  in the actual world the
measurement at $t_1$ has been performed and $\sigma_x(t_1)=1$ has been
obtained  does not ensure  that in a counterfactual world in which
$\sigma_x$ was not measured at $t_1$, but at a later time $t_2$, the
outcome has to be 
$\sigma_x(t_2)=1$, rather we still have probability $1\over 2$ for this
result. Thus, counterfactual reasoning leads us to the erroneous result
that the probability for $\sigma_x(t_1) =\sigma_x(t_2) =1$ is ${1\over
  2}\times {1\over 2} ={1\over 4}$.

\vskip 1.cm \noindent
{\bf 5. Time asymmetry prejudice.~~}
In my approach the pre- and post-selected states are given. Only
intermediate measurements are to be discussed. So the frequently posed
question about the probability of the result of the post-selection
measurement is irrelevant.  It seems to me that the critics of the
time-symmetrized quantum theory use in their arguments the
preconception of an asymmetry. It is not surprising then that they
reach various contradictions.  Probably the first to go according to
this line were Bub and Brown:
\begin{quotation}
Put simply, systems initially in the state $\psi_I$ which are subjected
to an $N$ measurement, and subsequently yield the state $\psi_F$ after
an $M_F$ measurement, would not necessarily yield this final state if
subjected to a measurement of $M$ instead of $N$. (1986, 2338)
\end{quotation}
Their argument is valid (see Albert et al. 1986) in
the context of the possible hidden variable theories which allow us to {\em
  predict} the results of measurements, but it should not be brought
against proposals of a time-symmetrized formalism as it was done
frequently later.  Let me consider two examples: Cohen  writes:
\begin{quotation}
We have no reason to expect that, for example, the $N/4$ systems
preselected by $|\psi_1(t_1)\rangle$ and post-selected by
$|\psi_2(t_2)\rangle$ after an intermediate measurement of
$\sigma_{1y}$ would still have yielded the state  $|\psi_2\rangle$
after an intermediate measurement of $\sigma_{2x}$ or of
$\sigma_{1y}\sigma_{2x}$ instead of $\sigma_{1y}$.(1995, 4375)    
\end{quotation}
A consistent time-symmetrized approach should question the pre-selection
on the same footing as the post-selection; or rather not question any of
them, as I propose.

Another asymmetry pre-conception leads to the ``retrodiction paradox''
of Peres (1994). The
asymmetry which he  considered  is, in fact, between {\em
  prediction} based on the results of measurements in the past of time $t$ in
question and {\em inference} based on the results of measurements
performed both in the past and
in the future of  time $t$. This inference he erroneously considered
as retrodiction (see Aharonov and Vaidman 1995).

If we are not considering a pre- and post-selected system then there
is an asymmetry between prediction and retrodiction.  For example
(Aharonov and Vaidman 1990, 11-12), assume that the $ x$ component of
the spin of a spin-${1\over 2}$ particle was measured at time $t$, and was
found to be $\sigma_x =1$.  While there is a symmetry regarding
prediction and retrodiction for the result of measuring $\sigma_x$
after or before time $t$ (in both cases we are certain that $\sigma_x
=1$), there is an asymmetry regarding the results of measuring
$\sigma_y$. We can predict equal probabilities for each outcome,
$\sigma_y  = \pm 1$, of a measurement performed after time $t$, but we
cannot claim the same for the result of a measurement of $\sigma_y$
performed before time $t$.  The difference arises from 
 the usual assumption that there is no
``boundary condition'' in the future, but there is a boundary
condition in the past: the state in which the particle was prepared
before time $t$. Maybe in a somewhat artificial way we can reconstruct
the symmetry even here, out of the context of pre- and post-selected
systems. We can ``erase'' the results of the measurements of the spin
measurements in the past (Vaidman 1987, 61). In order to do this we
perform at time $t_0$, before time $t$, a measurement of a Bell-type
operator on our particle and another auxiliary particle, an  ancilla.
We ensure that no measurement is performed on the particle between
$t_0$ and $t$ (except the possible measurement whose result we want to
consider) and we prevent any measurement on the ancilla  from
time $t_0$ and on. 
 The Bell-type measurement
correlates the quantum state of our particle evolving from the past
with the results of the future measurement performed on the ancilla.
Since the latter is unknown, we obtain, effectively, an unknown past for
our particle. Now, for such a system, if we know that the result of the
measurement at  time $t$ is $\sigma_x =1$, we can also {\em
  retrodict} that there are equal probabilities for both outcomes of
the measurement of $\sigma_y$ performed before time $t$ (but after
time $t_0$).  The time symmetry is restored.\footnote{The time symmetry
  is restored not just for the $\sigma_y$ measurement, but for any
  spin measurement.}

\vskip 1.cm \noindent
{\bf 6. Time symmetry of the process of measurement.~~}
Obviously, in order to discuss a measurement at time $t$ between two
other measurements in a time symmetric
way, the  process  of measurement at time $t$ 
 must  be time symmetric. Usually, a 
measurement of a quantum observable $A$  is modeled by the von Neumann
(1955) Hamiltonian 
\begin{equation}
  \label{neumann}
 H = g(t) p A,
\end{equation}
where $p$ is the momentum conjugate to the pointer variable $q$, and
the normalized coupling function $g(t)$ specifies the time of the
measurement interaction. The function $g(t)$ can be made symmetric in
time (not that it matters) and the form of the coupling then is
time symmetric. The result of the measurement is the difference
between the value of $q$ before and after the measurement interaction.
So it seems that everything is time symmetric.

However, usually there is an asymmetry in that that the initial position
of the pointer is customized to be zero (and therefore the final
position correspond to the measured value of $A$). This seemingly
minor aspect points  to a genuine asymmetry. Of course, the initial
zero position of the pointer  is  not
a necessary condition; we can choose any other initial position as well.
But, we cannot chose the
final position.   We {\em know} the initial position and we
find out, at the end of the measurement the final position. We can
introduce another step with symmetrical coupling, but we will not be
able to remove the basic asymmetry: we do not know the result of a
measurement before the measurement but we do know it after the
measurement.

This asymmetry in time is an intrinsic property of the concept of
measurement and is not specific  to the quantum theory.  It is
related to the arrow of time based on the increasing memory. See
illuminating discussion of Bitbol (1988) of the process of measurement
in the framework of the many-worlds interpretation  (Everett, 1957).

The symmetry aspect of the process of measurement which is important
for our discussion is that the process of measurement at time $t$
affects identically  forward
and backward evolving states.
A possible operational test of this condition is, that the
probabilities for     measurements performed immediately
after $t$,
 given a certain incoming state and  no information from the future, are identical to
 probabilities for  the same measurements performed immediately before
 $t$, given the same (complex conjugate) incoming state evolving backward in
 time and  no information from the past (see erasure procedure described
 in previous section).
The  measurement described by the
Hamiltonian (12) has this time symmetry.  Moreover, any ``ideal'' von
Neumann measurement, which projects on the property to be measured and
does not change the quantum state if it has the measured property, is
symmetric in this sense.

Recently, Shimony (1997) proposed considering more general quantum
measurements which do not change the measured property (so they are
repeatable, the main property which is required from a ``good''
measurement) but which change the state (even if it has the measured
property). He constructed an example of this kind of measurements for
which the ABL formula yields incorrect probabilities.  However, the
measurement he proposed is explicitly asymmetric in time and it breaks
the time symmetry requirement stated above.  Therefore, one should not
expect the ABL formula to hold for this case.\footnote{Note also an
  unusual property of Shimony's measurements in a situation in which
  the ABL formula is not involved: measurements of this type of
  commuting observables might disturb each other.}

In the appendix to his paper Shimony presented an example which shows
that ``The second kind of time symmetry, concerning the interchange of
initial and final conditions, fails even in the case of ideal
measurements.''  Indeed, this symmetry holds only (as is assumed in
this paper and in most papers on two-state vector formalism) if the
free Hamiltonian is zero. In Shimony's example the free Hamiltonian
was not zero, he assumed only that it is time independent. The basic
symmetry is for interchanging backward and forward evolving states
``entering'' time $t$, and thus the free Hamiltonian in the periods of
time $(t_1, t)$ and $(t, t_2)$ is irrelevant.  It is relevant only for
choosing appropriate measurements at $t_1$ and $t_2$ in order to get
this basic symmetry. This symmetry is
formally equivalent to ``the first kind of time symmetry of Shimony''
in which he considered ``reversal of measurement''. Reversed
measurement is related to reversed time's arrow and although it can be
analyzed theoretically we cannot make  Shimony's reversed measurements in
laboratory. This is why, instead of discussing ``reversed measurement''
at $t$, I prefer to discuss time symmetry under
appropriate interchange of measurements before and after time $t$.

\vskip 1.cm \noindent
{\bf 7. Elements of Reality.~~}
  Important counterfactuals in quantum
theory are  ``elements of reality''. For comparison, I'll start with a
definition of time asymmetric element of reality: 
 \begin{quotation}
 {\bf (v)~}   If we can {\it predict} with certainty that the result of
   measuring at time $t$ of an observable $A$ is $a$, then, at time
   $t$, there exists an element of reality $A=a$.
  \end{quotation}
  This is, essentially, a quotation from Redhead (1987), who, however
  considered it as a sufficient condition and not as a definition.
  Redhead was inspired by the criteria for elements of reality of
  Einstein, Podolsky and Rosen (EPR). In spite of similarity in its
  form, the EPR criteria, taken as a definition, is very different:
  ``If, without in any way disturbing the system, we can predict with
  certainty the value of a physical quantity...''  The crucial
  difference is that ``predict'' in the EPR definition means to find
  out using certain (non-disturbing) measurements, while in my
  definition ``predict'' means to deduce using existing information.
  Thus, for two spin-${1\over 2}$ particles in a singlet state, the
  value of a spin component of a single particle in any direction is
  an element of reality in the EPR sense (it can be found out by
  measuring another particle) and there is no elements of reality for
  a spin component value in any direction according to my definition
  (in the EPR state, the probability to find spin ``up'' in any direction
  is ${1\over 2}$).

 Definition (v) of elements of reality is
  asymmetric in time because of the word ``predict''. I have proposed
  a modification of this definition applicable for time symmetric
  elements of reality (Vaidman 1993a):
 \begin{quotation}
 {\bf (vi)~}   If we can {\em infer} with certainty that the result of
   measuring at time $t$ of an observable $A$ is $a$, then, at time
   $t$, there exists an element of reality $A=a$.
\end{quotation}
The word ``infer'' is neutral relative to past and future. The
inference about results at time $t$ is based on the results of
measurements on the system performed both before and after time $t$.
Note, that in some situations we can ``infer'' more facts than can be
obtained by ``prediction'' based on the results in the past and
``retrodiction'' based on the results in the future (relative to $t$)
together.

The difference between definitions of ``elements of reality'', (v) and
(vi), and definitions of counterfactuals in quantum theory (iv) and
(iii) is that the property ${\cal P}$ in (v) and (vi) is constrained
to ``the result of measuring at time $t$ of an observable $A$ is
$a$''.  In fact, time asymmetric ``elements or reality'' (v), defined
as ``predictions'', do not represent ``interesting'' counterfactuals.
There is no nontrivial set of such counterfactual statements, i.e.,
set of statements which cannot be tested all on a single system.
Indeed, all observables the measurement of which yield definite
outcomes for a pre-selected system can be tested together.  One way to
extend the definition of time asymmetric elements of reality in order
to get nontrivial counterfactuals is to consider ``multiple-time
measurements'' (instead of measurements at time $t$ only). Another
extension, which corresponds to numerous analysis in the literature,
is to go beyond statements about observables which have definite
values:
 \begin{quotation}
  {\bf (vii)~}    If we can {\it predict} with certainty a certain relation
   between  the results $a_j$ of
   measuring at time $t$ a set of observables $A_j$, then, at time
   $t$, there exists a ``generalized element of reality'' which is this
   relation between $a_j$'s.
  \end{quotation}

A simple example of this kind is a system of  two spin-${1\over 2}$
particles prepared, at
$t_1$, in a singlet
state 
\begin{equation}
|\Psi_1\rangle = {1\over {\sqrt 2}}(  |{{\uparrow}}\rangle_1
|{\downarrow}\rangle_2 -   |{\downarrow}\rangle_1
|{{\uparrow}}\rangle_2).
\end{equation}
 We can predict with certainty that 
the results of measurements of spin components of the two particles
fulfill the following two relations:
\begin{eqnarray} 
\{\sigma_{1x}\} + \{\sigma_{2x}\} = 0 , \\
\{\sigma_{1y}\} +\{ \sigma_{2y}\} = 0 ,
\end{eqnarray} 
where $\{\sigma_{1x}\}$ signifies the result of measurement of spin $x$
component of the first particle, etc.  The relations (14) and (15)
represent set of generalized elements of reality (vii). This is a
nontrivial set of counterfactuals because (14) and (15) 
cannot be tested together: the measurement of $\sigma_{1x}$
disturbs the measurement of $\sigma_{1y}$ as well as the measurement
of $\sigma_{2x}$ disturbs the measurement of $\sigma_{2y}$.

In contrast,   the set of elements of reality  
(vi) given by 
\begin{eqnarray} 
\{\sigma_{1x} + \sigma_{2x}\} = 0 , \\
\{\sigma_{1y} + \sigma_{2y}\} = 0 ,
\end{eqnarray} 
can be tested on a single system, see
  Aharonov and Vaidman (1986) for description of such measurements.
Yet another set of counterfactuals, which consists of definite
statements about measurements, but which does not fall
into category (vi) because these are {\em two-time}  measurements performed at two
different time moments $t_1$ and $t_2$, {\em cannot} be tested on a
single system:
\begin{eqnarray} 
\{\sigma_{1x}(t_1) + \sigma_{2x}(t_2)\}  = 0 , \\
\{\sigma_{1y}(t_2) + \sigma_{2y}(t_1)\} = 0 .
\end{eqnarray} 
Note a  situation which involves only a single free
spin-${1\over 2}$ particle. The particle is  prepared, before $t_1$, $t_1 < t_2 < t_3$, in the state
$|{{\uparrow}}_y\rangle$. Then, a nontrivial set of counterfactuals is:
\begin{eqnarray} 
\{\sigma_{x}(t_1) - \sigma_{x}(t_3)\} = 0 , \\
\{\sigma_{y}(t_2)\}  = 1 .
\end{eqnarray} 
 In this example, however, statement (20) has somewhat different
character because  
it  depends not   on the results of measurements performed on the
particle before or after the period of time $(t_1, t_2)$, but on the
fact that the system was not disturbed during this period of time.

\vskip 1.cm \noindent
{\bf 8. Quantum Puzzles.~~}
In this section I shall describe a few interesting tasks which cannot
be performed in the framework of classical physics and in the solution of
which quantum counterfactuals are involved.

The first example considers a game for a team of three players, $A, B$
and $C$.  The players are allowed to make any preparations before 
they  will be taken to three remote locations. Then, at a
certain time, each of them will be asked one out of two possible questions:
``what is the value of $X$?'' or ``what is the value of $Y$?''.
According to the rules of the game, either all three will be asked the
$X$ question, or just one will be asked the $X$ question and the two
others will be asked the $Y$ question. The answers to both questions
are limited to two possibilities: $1$ or $-1$.  In order to win, the
answers of the team should fulfill one of the following relations:
 \begin{eqnarray}
X_A X_B X_C = -1 ,\\
X_A Y_B Y_C = 1 ,\\
Y_A X_B Y_C = 1 ,\\
Y_A Y_B X_C = 1 ,
\end{eqnarray}
with obvious notation: $X_A$ is the answer of $A$ on  question $X$,
etc. What should the team do in order to win for sure?

It seems that that the task is impossible. Since the distance between
the players of the team does not allow interchange of any messages
sent after they were asked the questions, it is natural to assume that
delaying the decision  of which answer to give for each question until the
question is actually asked cannot help. Thus, an optimal strategy
should correspond to definite decisions of each player which answers
to give for both possible questions. But it is easy to prove that any such
strategy cannot ensure winning for all allowed combinations of
questions. Indeed, if it does, then there must be a set of answers
$\{X_A, Y_A, X_B, Y_B, X_C, Y_C\}$ such that all equations (22-25) are
fulfilled. This however is impossible, because the product of all left
hand sides of equations (22-25) is the product of squares of numbers
which are $\pm 1$ and therefore it equals 1, while the product of all right
hand sides of these equation yields $-1$.

Nevertheless, quantum theory  provides a solution (Greenberg, Horne, and
Zeilinger 1989;  Mermin 1990). Each member of the team takes with him a
spin-${1\over 2}$ particle. The particles are prepared in a correlated  state 
\begin{equation}
|\Psi\rangle = {1\over \sqrt
  2}(|{\uparrow}\rangle_A|{\uparrow}\rangle_B|{\uparrow}\rangle_C - 
|{\downarrow}\rangle_A|{\downarrow}\rangle_B|{\downarrow}\rangle_C) . 
\end{equation}
Now, if a member of the team is asked  the $X$ question, he measures
$\sigma_x$ and gives the answer which he obtains in his
experiment. If he is asked the $Y$ question, he measures $\sigma_y$
instead. Quantum theory ensures that the team  following  this strategy 
{\em always} wins.

For the system of three spin-${1\over 2}$ particles in the state (26) we have the
following set of  ``generalized elements of reality'' (vii):
  \begin{eqnarray}
\{{\sigma_A}_x\} \{{\sigma_B}_x\} \{{\sigma_C}_x\} = -1 ,\\
\{{\sigma_A}_x\} \{{\sigma_B}_y\} \{{\sigma_C}_y\} = 1 ,\\
\{{\sigma_A}_y\} \{{\sigma_B}_x\} \{{\sigma_C}_y\} = 1 , \\
\{{\sigma_A}_y\} \{{\sigma_B}_y\} \{{\sigma_C}_x\} = 1 .
\end{eqnarray}
Here $\{{\sigma_A}_x\}$ signifies the outcome of the measurement  of
$\sigma_x$ by  player $A$ etc. 
 There is no possibility
of  having a system of three particles on which all these relations are tested.
Therefore, this is a nontrivial set of counterfactuals.

The importance of this example is not only that it shows how a
``quantum'' team can  win against any ``classical'' team. This example yields a
simple proof of the nonexistence  of local hidden variables 
telling  the results of each local measurement prior to their
performance.

Let us turn now  to a puzzle which involves quantum time-symmetrized
counterfactuals (Vaidman et al. 1987, Mermin 1995).
Alice and Bob play the following game: Alice has a spin-${1\over 2}$ particle
which she gives  to Bob at time $t$ when he can perform a single spin
component measurement in the $\hat x$, $\hat y$, or $\hat z$ direction. After the
measurement, Bob gives the particle back to Alice without telling her
which measurement he did. Alice's task is to tell Bob  the
results of his measurements for all his possible choices. 

In this problem Alice makes a  measurement before time $t$ and
another measurement after time $t$. At the end she must know a
nontrivial set of counterfactual statements about measurements at
time $t$:
  \begin{eqnarray}
\{{\sigma_x}\} = x ,\\
\{{\sigma_y}\} = y ,\\
\{{\sigma_z}\} = z ,
\end{eqnarray}
where $x, y$ and $z$ can have values $\pm 1$. Note that (31-33) are
elements of reality according to  definition (vi).

Let us first consider a trivial puzzle of this kind, when Bob is
limited to measurements of spin  components in only two directions,
$\hat x$ and $\hat y$. All what
Alice has to do in this case is to measure, say, $\sigma_x$ before
time $t$ and $\sigma_y$ after time $t$. Then, the result of the first
measurement is $x$ and the second measurement yields $y$.

If Bob has a choice of three possible spin
measurements, the solution is much more difficult. Alice should
prepare a correlated state of this and another spin-${1\over 2}$
particle, keep the other particle undisturbed, and make a special joint
measurement on both particles after she gets the first one from Bob
(see details in Vaidman et al. 1987). She cannot control the outcome
of her second measurement, but she can construct the measurement in
such a way that for every outcome (out of four possibilities) she will
have an infinite set of time symmetric counterfactuals about
measurements at time $t$: spin components for every direction
belonging to a certain cone will be known. For appropriate measurement
all four possible cones will include $\hat x$, $\hat y$, and $\hat z$
axes. Thus, for any possible outcome of the final measurement Alice
will know the set of counterfactuals (31-33).

Obviously, the set of statements (31-33) is a ``nontrivial'' set of
counterfactuals. Indeed, elements of reality (31-33) cannot be tested
on a single particle because the operators $\hat\sigma_{x}$,
$\hat\sigma_{y}$ and $\hat\sigma_{z}$ do not commute.  It is
interesting that in some cases, elements of reality of a pre- and
post-selected quantum system related even to commuting observables
cannot be tested together. Consider again two spin-${1\over 2}$ particles
prepared in the singlet state (13). At $t_2$ the particles are found in the
state $|\Psi_2\rangle = |{\uparrow}_x\rangle_1 |{\uparrow}_y\rangle_2$.  A  set of elements of reality for these particles at an
intermediate time is (use the ABL formula (2) to see this):
\begin{eqnarray} 
\{{\sigma_1}_y\} = -1 , \\
\{{\sigma_2}_x\} = -1 ,\\
\{{\sigma_1}_y {\sigma_2}_x\}=-1 .
\end{eqnarray}
Even the first two elements of reality (34-35) cannot be tested
together, in spite of the fact that these are statements about  two separated
particles.

Another interesting feature of time symmetric  elements of reality
(vi) of a pre- and  post-selected quantum system is that the
``product rule'' does not hold. The product rule means that  if $A=a$ and $B=b$ are elements
of reality, then $AB =ab$ is also an  element of reality. (This
product rule does hold for elements of reality (v) of a pre-selected
quantum system.)   Indeed, in the above example (34-36) we have:
$\{{\sigma_1}_y {\sigma_2}_x\} \neq \{{\sigma_1}_y\}
\{{\sigma_2}_x\}$. Note, that in this example there is no ``generalized
element of reality''  (vii) for $\{{\sigma_1}_y\} \{
{\sigma_2}_x\}$: the product of the results of simultaneous measurements of
observables 
${\sigma_1}_y$ and ${\sigma_2}_x$ can be 1 or $-1$. Therefore, the
product rule of a different type: ``if $A=a$ and $B=b$ are elements of
reality, then $\{A\}\{B\} =ab$ is a generalized element of reality''
does not hold either.

The failure of the product rule is important for discussing Lorentz
invariance of a realistic quantum theory. Recently,  extensive
discussions of the impossibility of a realistic Lorentz invariant quantum
theory were based on the (unjustified) assumption of the validity of the
product rule (Hardy 1992; Vaidman 1993a, 1993b; Cohen and Hiley 1995, 1996).

\vskip 1.cm \noindent
{\bf 9. Weak Measurements.~~}
One might argue about the significance of time-symmetrized
counterfactuals  beyond the
philosophical construction of (jointly unmeasurable) ``elements of
reality'', since it is impossible to perform incompatible measurements
on a single system. I find the most important aspect of these concepts
to be  their relation to {\em weak measurements} (Aharonov and Vaidman
1990).

 Weak measurements are almost standard
measurement procedures with weakened coupling. Weak measurements
essentially do not change the quantum states (evolving forward and
backwards in time) of the system.  Several weak measurements can be
performed on a single system and they are compatible even though their
counterparts, the ideal measurements are not compatible. The outcomes
of weak measurements are {\em weak values}, Eq. 3. Weak values have
many interesting properties, in particular $(A+B)_w= A_w+ B_w$ even
for non-commuting observables $A$ and $B$.  I also
defined the outcomes of weak measurements as {\em weak-measurement elements of reality} (Vaidman,
1996).  The weak value
is not just a theoretical concept related to a gedanken experiment.
Recently, weak values have been measured in a real laboratory (e.g. Ritchie
et al. 1991).

An  interesting connection between weak and strong
(ideal) measurements is the theorem (Aharonov and Vaidman 1991) which
says that if the probability for a certain value to be the result of a
strong measurement is 1, then the corresponding weak measurement must
yield the same value, i.e. element of reality $A=a$, implies
weak-measurement element of reality $A_w=a$. 
Although a  set of statements about weak-measurement elements of reality does
not fall under the category of ``a nontrivial set of counterfactuals''
(because one can always consider a pre- and post-selected ensemble on
which all week measurements of the set are performed
together),  I find  the application of ``counterfactuals'' for
such  situations  most appropriate.  When we discuss the outcomes of weak
measurements, the statements about strong measurement are literally
counterfactual.  The reasoning about {\it unperformed} strong
measurements helps us to find out the results of {\em performed} weak
measurements. For example, in the situation described by elements of
reality (31-33)  the weak measurement of $ \sigma_{x} +
\sigma_{y}+\sigma_{z}$ will yield $ (\sigma_{x} + \sigma_{y}  +\sigma_{z})_w= (\sigma_{x})_w
+(\sigma_{y})_w +(\sigma_{z})_w  = x + y +z$. It might be that
$x=y=z=1$. Then,  $ (\sigma_{x} + \sigma_{y}  +\sigma_{z})_w =3$,
 in spite of the fact that the
eigenvalues of  $ \sigma_{x} +
\sigma_{y} +\sigma_{z}$ are $\pm \sqrt 3$.  Moreover,  a weak measurement of spin in any direction will yield a
projection of a ``weak-measurement spin vector'' whose size is $
3$  on this direction (Aharonov and Vaidman 1991).

 Let me add another, even more striking,
example of a situation in which counterfactuals about strong
measurements help to find out properties of weak measurements.
Consider a single particle prepared at $t_1$ in a superposition of being in
three separated boxes:

\begin{equation}
|\Psi_1\rangle = {{1\over \sqrt
  3}}(|A\rangle + |B\rangle + |C\rangle) .
\end{equation}
At later time $t_2$ the particle was found in another superposition:

\begin{equation}
|\Psi_2\rangle = {{1\over \sqrt
  3}}(|A\rangle + |B\rangle - |C\rangle).
\end{equation}
In between no measurement was performed on this particle, but it
coupled very weekly to some other systems. The question is: how one can
characterize this coupling?

A set of counterfactual statements for this particle is 
\begin{eqnarray} 
{\rm \bf P}_A = 1 ,\\
{\rm \bf P}_B = 1 ,\\
{\rm \bf P}_A + {\rm \bf P}_B + {\rm \bf P}_C = 1 .
\end{eqnarray}
Or, in words: if we open box $A$, we find the particle there for sure;
if we open box $B$ (instead), we also find the particle there for sure; if we
open all boxes, we find the particle there for sure. These
counterfactual statements help us to find out statements about
weak-measurement elements of reality: 
\begin{eqnarray} 
({\rm \bf P}_A)_w = 1 ,\\
({\rm \bf P}_B)_w = 1 ,\\
({\rm \bf P}_A + {\rm \bf P}_B + {\rm \bf P}_C)_w = 1 .
\end{eqnarray}
From these results we can also deduce that $({\rm \bf P}_C)_w = -1$. Thus, the
counterfactual statements (39-41) help us to answer the question posed
above: for every sufficiently weak interaction, the effective coupling
to this single particle is equivalent to the weak coupling to a single
particle in box $A$, a single particle in box $B$ and {\em minus} one
particle in box $C$. The meaning of the latter is that for a pre- and
post-selected ensemble of many such particles there is an effective
{\em negative} pressure in box $C$. (An experiment which will test
this is very difficult because the probability to obtain such an ensemble
is very low.)

\vskip 1.cm \noindent
{\bf 9. Conclusions.~~}
The main goal of this paper is to defend the  time-symmetrized quantum
formalism, originated by the ABL paper,  against recent
criticism.
The time-symmetrized formalism of quantum theory requires certain
properties of quantum measurements. I discussed some proposed time
asymmetric procedures which fulfill some of the properties of quantum
measurement, and for which the time-symmetrized formalism yields
incorrect predictions. I showed, however, that for any ideal
measurement as well as for any von Neumann type measurement (ideal or not
ideal) the time-symmetrized formalism is valid.  This answers some of
the criticism of the time-symmetrized formalism.
 However, the key argument
of the critics was different; it was  an alleged proof, repeated by several
authors, showing that in certain situations counterfactual
interpretation of the ABL rule leads to a contradiction with the predictions
of quantum theory.
I argued here that these situations  were not
of ``counterfactual'' nature. The way of reasoning which lead the
authors to a contradiction was not the ``counterfactual interpretation of
the ABL rule'' but a logical error. The error was in calculating
the probability for the result of a certain experiment assuming that this
experiments was not  performed. Not only that the authors of the proof
assume that the measurement was not performed in the actual world (as
is usually done in discussing counterfactuals in quantum theory) but
also that it was not performed in the counterfactual world, for which
the statements about the probability of the result of this measurement was
made.

The argumentation of the critics of the time-symmetrized quantum
formalism was that the ABL rule for the cases when the measurement was
actually performed is correct, but philosophically not interesting,
while the counterfactual interpretation is interesting but incorrect.
Showing that their ``counterfactual interpretation'' is incorrect
still leaves us with the claim that time-symmetrized formalism
``yields no fresh insights about the fundamental interpretive issues
in quantum mechanics'' (Sharp and Shanks, 1993).  In order to refute
this argument I made an analysis of counterfactuals in quantum theory
which I believe is important by itself.  I narrowed the concept of
``counterfactuals'' to statements about properties of the results of
quantum measurements which would be valid if these measurements were
performed.  The crucial issue for counterfactuals in quantum theory is
what is held ``fixed'' in counterfactual worlds. I proposed defining
``results of all measurements (except measurements at time $t$)'' as
all that is fixed.  I showed that this definition for a pre-selected
situations is equivalent to the usual definition of counterfactuals.
However, contrary to the usual approach, this definition is also
applicable for pre- and post-selected situation, and it yields
time-symmetrized counterfactuals with the results of measurements
fixed both in the past and in the future.

The first nontrivial  type of situations for which these counterfactuals can be
applied is when there is a set of  statements which cannot be tested
on a single system. Therefore, the set of  counterfactual
  statements about quantum system on which these measurement might have
  been performed cannot be  equivalent to a set of statements about
  actually  performed measurements. There are such sets of
  nontrivial counterfactuals  both for pre-selected only  quantum systems and
for   pre- and post-quantum systems. (Note, however, that only for
pre- and post-selected situations there are nontrivial sets of
counterfactuals which are ``elements of reality'', i.e. definite
results of measurements. For pre-selected situations,  definite
{\em properties} of the results of measurements have to be considered.)
These nontrivial sets of counterfactuals play an important role in
ongoing discussions of the locality and Lorentz invariance of quantum
theory.

The second application of time-symmetrized counterfactuals I have presented
here is the connection between ``elements of reality'' and weak coupling to a
quantum system. I considered a pre- and post-selected system on which
no intermediate strong measurements were performed. Then, ``elements
of reality'' are literally counterfactual statements about definite
results of measurements which actually were not performed. These
counterfactual statements helped to characterize weak measurements  which actually took place. 
The outcomes of these  weak measurements are  weak values (another
time symmetric concept) which yield a  simple  picture
of a pre- and post-selected quantum system. I find this picture
important because it 
allowed us to see numerous surprising quantum effects which are
hidden in the framework of the standard approach in  a very complicated
mathematics of some peculiar interference effects (e.g. Aharonov et
al. 1987, 1990, 1993; Vaidman 1991).

It is a pleasure to thank Yakir Aharonov, David Albert, Avshalom
Elitzur, Lior Goldenberg, Yoav Ben-Dov, Igal Kvart, Abner Shimony and
Stephen Wiesner for helpful discussions and Willy De Baere, Niall
Shanks, and David Sharp for useful correspondence.  The research was
supported in part by grant 614/95 of the Israel Science Foundation.

\vskip .8cm
\vfill
\break

 \centerline{\bf REFERENCES}
\vskip .15cm
\footnotesize

\vskip .13cm \noindent 
Aharonov, Y. and Albert, D. (1987), 
``The Issue of Retrodiction in Bohm's Theory'',
in B.J. Hiley and F.D. Peat (eds.) {\em Quantum Implications}. New
York: Routledge \& Kegan Paul, pp.224-226.

\vskip .13cm \noindent 
 Aharonov, Y., Albert, D., Casher, A., and Vaidman, L. (1987),
 ``Surprising Quantum Effects'',
{\em  Physics Letters A 124}: 199-203.

\vskip .13cm \noindent 
Aharonov, Y.,  Albert, D., and D'Amato, S. (1985),
``Multiple-Time Properties of Quantum Mechanical Systems'',
 {\em Physical Review  D 32}:  1975-1984.

\vskip .13cm \noindent 
Aharonov, Y., Anandan, J., Popescu, S., and Vaidman, L.  (1990),
``Superpositions of Time Evolutions of a Quantum System and a Quantum Time
Machine'',
{\em  Physical Review Letters 64}: 2965-2968.
 
\vskip .13cm \noindent 
Aharonov, Y.,  Bergmann,  P.G., and  Lebowitz, J.L. (1964),
``Time Symmetry in the Quantum Process of Measurement'',
 {\em Physical Review 134B}: 1410-1416. 

\vskip .13cm \noindent 
Aharonov, Y.,  Popescu, S., Rohrlich, D., and Vaidman, L.  (1993),
 ``Measurements, Errors, and Negative Kinetic Energy'',
 {\em Physical   Review A 48}: 4084-4090.
 
\vskip .13cm \noindent 
Aharonov, Y. and Vaidman, L. (1990),
 ``Properties of a Quantum System
During the Time Interval Between Two Measurements'',
{\em Physical Review A 41}: 11-20.

\vskip .13cm \noindent 
Aharonov, Y. and Vaidman, L. (1991),
``Complete Description of a Quantum System at a Given Time'',
{\em Journal of  Physics  A 24}: 2315-2328.

\vskip .13cm \noindent 
Aharonov, Y. and Vaidman, L. (1995),
``Comment on `Time Asymmetry in Quantum Mechanics: a Retrodiction Paradox' '',
{\em Physical Letters A 203}: 148-149.

\vskip .13cm \noindent 
Albert, D.,  Aharonov, Y. and  D'Amato, S. (1986),
``Comment on `Curious Properties of Quantum Ensembles which have been Both
  Preselected and Post-Selected' '',
{\em Physical Review Letters 56}: 2427.

\vskip .13cm \noindent 
Bedford, D. and Stapp, H.P. (1995),
``Bell's Theorem in an Indeterministic Universe'',
{\em Synthese 102}: 139-164.

\vskip .13cm \noindent 
Bennett, J. (1984),
``Counterfactuals and Temporal Direction'',
{\em Philosophical Review 93}: 57-91.

\vskip .13cm \noindent 
Bitbol, M. (1988),
``The Concept of Measurement and  Time-Symmetry in  Quantum Mechanics'',
{\em Philosophy of Science 55}: 349-375.

\vskip .13cm \noindent 
 Bub, J. and Brown, H. (1986), 
``Curious Properties of Quantum Ensembles which have been Both
  Pre-Selected and Post-Selected'',
{\em Physical Review Letters 56}: 2337-2340.

\vskip .13cm \noindent 
Cohen,  O. (1995), 
``Pre- and Post-Selected Quantum Systems, Counterfactual Measurements,
and Consistent Histories'',
 {\em Physical Review  A 51}: 4373-4380. 
 
\vskip .13cm \noindent 
 Cohen, O. and Hiley, B.J. (1995),
  ``Reexamining the Assumptions that Elements of Reality can be
  Lorentz Invariant'', {\em Physical Review A 52}: 76-81.

\vskip .13cm \noindent 
 Cohen, O. and Hiley, B.J. (1996), 
``Elements of Reality, Lorentz Invariance and the Product Rule'',
 {\em Foundations of Physics 26}: 1-15.

\vskip .13cm \noindent 
Everett, H. (1957),
 `` `Relative State' Formulation of Quantum Mechanics'',
{\em Review of Modern Physics 29}: 454-462.

\vskip .13cm \noindent 
Greenberger, D.M., Horne, M.A. and Zeilinger, A. (1989),
``Going  Beyond Bell's Theorem'',
in Kafatos, M. ed. {\em Bell Theorem, Quantum Theory and Conceptions
  of the Universe}. Kluwer, Dordrdecht pp.69-72.

\vskip .13cm \noindent 
Hardy, L. (1992)
``Quantum Mechanics, Local Realistic Theories, and Lorentz-Invariant Realistic Theories'',
{\em Physical Review Letters 68}: 2981-2984.

\vskip .13cm \noindent 
Lewis, D. (1973),
{\em Counterfactuals}.
Oxford:Blackwell Press.

\vskip .13cm \noindent 
 Lewis, D. (1986),
 ``Counterfactual Dependence and
  Time's Arrow'' reprinted from {\em Nous 13}: 455-476 (1979) and {\em
    Postscripts to ``Counterfactual ..''} in Lewis, D. {\em
    Philosophical Papers Vol.II}, Oxford: Oxford University Press,
  pp.32-64.

\vskip .13cm \noindent 
Mermin, N.D. (1989),
``Can You Help Your Team Tonight by Watching on TV? More Experimental
Metaphysics from Einstein, Podolsky, and Rosen'',
in J.T. Cushing and E. McMullin (eds.) {\em Philosophical Consequences
  of Quantum Theory: Reflections on Bell's Theorem}. Notre Dame:
University of Notre Dame Press, pp.38-59.

\vskip .13cm \noindent 
Mermin, N.D. (1990),
``Quantum Mysteries Revisited'',
{\em American Journal of Physics 58}: 731-734.

\vskip .13cm \noindent 
Mermin, N.D. (1995),
``Limits to Quantum Mechanics as a Source of Magic Tricks: Retrodiction
and the Bell-Kochen-Specker Theorem'',
{\em  Physical Review Letters 74}: 831-834.

\vskip .13cm \noindent 
Miller, D. J. (1996),
``Realism and Time Symmetry in Quantum Mechanics'',
{\em Physical Letters A 222}: 31-36.

\vskip .13cm \noindent 
Penrose, R. (1994),
{\em Shadows of the Mind}.
 Oxford: Oxford University Press.

\vskip .13cm \noindent 
Peres, A. (1978),
``Unperformed Experiments Have no Results'',
{\em American Journal of Physics 46}: 745-747.

\vskip .13cm \noindent 
Peres, A. (1993),
{\em Quantum Theory: Concepts and Methods}.
Dordrecht: Kluwer Academic Publisher.

\vskip .13cm \noindent 
Peres, A. (1994),
``Time Asymmetry in Quantum Mechanics: a Retrodiction Paradox'',
{\em Physical Letters A 194}: 21-25.
 
\vskip .13cm \noindent 
Price H. (1996),
{\em Time's Arrow \& Archimedes' Point}. New York: Oxford University Press.

\vskip .13cm \noindent 
Redhead, M. (1987),
{\em Incompleteness, Nonlocality, and Realism: a Prolegomenon to the
  Philosophy  of Quantum Mechanics}. New York: Oxford University Press.

\vskip .13cm \noindent 
Ritchie, N.W.M.,  Story, J.G. and  Hulet, R.G. (1991),
``Realization of a `Measurement of a Weak Value''',
{\em Physical Review Letters 66}: 1107-1110.

\vskip .13cm \noindent 
Sharp, W.D. and Shanks, N. (1993),
``The Rise and Fall of Time-Symmetrized Quantum Mechanics'',
{\em Philosophy of Science 60}: 488-499.

\vskip .13cm \noindent 
 Shimony, A. (1997),
``A Bayesian Examination of Time-Symmetry in the Process of
Measurement'', 
{\em Erkenntnis 45}: 337-348.

\vskip .13cm \noindent 
 Skyrms, B. (1982),
 ``Counterfactual Definetness and Local
 Causation'', {\em Philosophy of Science 49}: 43-50.

\vskip .13cm \noindent 
Vaidman, L. (1987),
``The Problem of the Interpretation of Relativistic Quantum
Theories'',
{\em Ph.D. Thesis}, Tel-Aviv University.

\vskip .13cm \noindent 
 Vaidman, L., Aharonov, Y., Albert, D., (1987),
``How to Ascertain the Values of $\sigma_x, \sigma_y,$ and $\sigma_z$ of
a Spin-${1\over 2}$ Particle'',
{\em  Physical Review Letters 58}: 1385-1387.

\vskip .13cm \noindent 
 Vaidman, L. (1991),
``A Quantum Time Machine'',
{\em Foundations of  Physics  21}: 947-958.

\vskip .13cm \noindent 
 Vaidman, L. (1993a),
``Lorentz-Invariant `Elements of Reality' and the Joint
Measurability of Commuting Observables'',
{\em Physical Review Letters 70}: 3369-3372.

\vskip .13cm \noindent 
Vaidman, L. (1993b),
`` `Elements of Reality' and the Failure of the Product Rule''
  in P.J. Lahti, P.~Bush, and P. Mittelstaedt (eds.),{\em Symposium on the Foundations  of Modern Physics}. New Jersey: World
Scientific, pp. 406-417.

\vskip .13cm \noindent 
Vaidman, L. (1996),
``Weak-Measurement Elements of Reality'',
 {\em Foundations of Physics 26}: 895-906.

\vskip .13cm \noindent 
Vaidman, L. (1997),
``On the validity of the Aharonov-Bergmann-Lebowitz rule'',
Tel-Aviv University preprint quant-ph/9703001.

\vskip .13cm \noindent 
von Neumann, J. (1955),
  {\em Mathematical Foundations of Quantum Theory}.
  Princeton: Princeton University Press.
 
\end{document}